\documentclass[12pt,a4paper]{iopart}
\usepackage{setstack,iopams}
\usepackage{graphicx} 
\usepackage[caption=false]{subfig}
\begin{document}


\newcommand{\braket}[2]{{\left\langle #1 \middle| #2 \right\rangle}}
\newcommand{\bra}[1]{{\left\langle #1 \right|}}
\newcommand{\ket}[1]{{\left| #1 \right\rangle}}
\newcommand{\ketbra}[2]{{\left| #1 \middle\rangle \middle \langle #2 \right|}}
\newcommand{\binom}[2]{{#1 \choose #2}}


\title{Quantum Walk Search on Johnson Graphs}

\author{Thomas G Wong}
\address{Faculty of Computing, University of Latvia, Rai\c{n}a bulv.~19, R\=\i ga, LV-1586, Latvia}
\ead{\mailto{twong@lu.lv}}

\begin{abstract}
	The Johnson graph $J(n,k)$ is defined by $n$ symbols, where vertices are $k$-element subsets of the symbols, and vertices are adjacent if they differ in exactly one symbol. In particular, $J(n,1)$ is the complete graph $K_n$, and $J(n,2)$ is the strongly regular triangular graph $T_n$, both of which are known to support fast spatial search by continuous-time quantum walk. In this paper, we prove that $J(n,3)$, which is the $n$-tetrahedral graph, also supports fast search. In the process, we show that a change of basis is needed for degenerate perturbation theory to accurately describe the dynamics. This method can also be applied to general Johnson graphs $J(n,k)$ with fixed $k$.
\end{abstract}

\pacs{03.67.Ac, 05.40.Fb, 02.10.Ox}


\section{Introduction}

\begin{figure}
\begin{center}
	\subfloat[] {
		\includegraphics{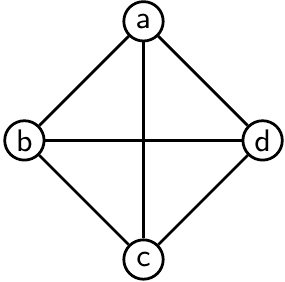}
		\label{fig:complete4}
	} \quad
	\subfloat[] {
		\includegraphics{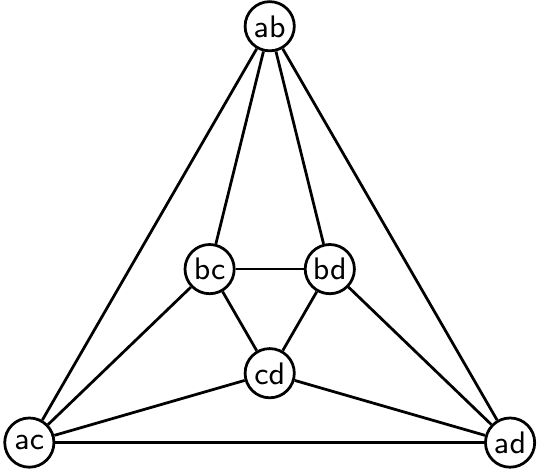}
		\label{fig:triangular4}
	}
	\caption{(a) The Johnson graph $J(4,1)$, which is the complete graph $K_4$. (b) The Johnson graph $J(4,2)$, which is the triangular graph $T_4$.}
\end{center}
\end{figure}

The Johnson graph $J(n,k)$ is defined by $n$ symbols, where vertices are $k$-element subsets of the symbols, and vertices are adjacent if they differ in exactly one symbol (Section 8 of \cite{HS1993}). Consider two examples with $n = 4$ symbols $\{a,b,c,d\}$. First, if $k = 1$, then the vertices of $J(4,1)$ are $1$-element subsets of the symbols, so the vertices are simply $a$, $b$, $c$, and $d$. Since they all differ from each other by one symbol, all the vertices are adjacent to each other, and the resulting graph is the complete graph $K_4$ shown in \fref{fig:complete4}. In general, $J(n,1)$ is the complete graph $K_n$. For the second example, if $k = 2$, then the vertices of $J(4,2)$ are $2$-element subsets of $\{a,b,c,d\}$, so the vertices are $ab$, $ac$, $ad$, $bc$, $bd$, and $cd$. Vertices are adjacent if they differ in exactly one element; for example, $ab$ and $ac$ are adjacent, while $ab$ and $cd$ are not. The resulting graph is shown in \fref{fig:triangular4}, and it is the triangular graph $T_4$. In general, $J(n,2)$ with $n \ge 4$ is the triangular graph $T_n$ \cite{Cameron1991}.

Johnson graphs have made recent headlines in the computer science and quantum computing communities \cite{Aaronson2015,Kun2015} due to Babai's quasipolynomial algorithm for graph isomorphism \cite{Babai2015}, in which the algorithm is polynomial in all cases except for Johnson graphs. That is, Johnson graphs are the only structures preventing the graph isomorphishm problem from being polynomial rather than quasipolynomial.

Despite this recent notoriety, Johnson graphs are familiar structures on which to perform quantum walks, with Ambainis's element distinctness algorithm \cite{Ambainis2004} famously walking on them in discrete-time. More recently, a continuous-time quantum walk algorithm for element distinctness was developed by Childs \cite{Childs2010}, which walks on a modified Johnson graph with additional vertices and edges acting as the oracle.

In this paper, we investigate \emph{spatial search} by continuous-time quantum walk \cite{CG2004} on Johnson graphs for a unique marked vertex. The system $\ket{\psi(t)}$ begins in an equal superposition $\ket{s}$ over the vertices:
\begin{equation}
	\label{eq:s}
	\ket{\psi(0)} = \ket{s} = \frac{1}{\sqrt{N}} \sum_{v=1}^N \ket{v},
\end{equation}
where
\begin{equation}
	\label{eq:N}
	N = \binom{n}{k} = \frac{n!}{(n-k)!k!}
\end{equation}
is the number of vertices of $J(n,k)$ since we choose $k$ symbols out of $n$ for each vertex. From this starting state, the system evolves by Schr\"odinger's equation
\[ \rmi \frac{\rmd}{\rmd t} \ket{\psi(t)} = H \ket{\psi(t)} \]
with Hamiltonian
\begin{equation}
	\label{eq:H}
	H = -\gamma A - \ketbra{w}{w}.
\end{equation}
Here, $\gamma$ is a real and positive parameter corresponding to the jumping rate (amplitude per time) of the quantum walk, $A$ is the adjacency matrix of the graph ($A_{ij} = 1$ if two vertices are adjacent and $0$ otherwise), and $\ket{w}$ is the marked vertex to search for (so $\ketbra{w}{w}$ acts as a Hamiltonian oracle \cite{Mochon2007}). Note that Johnson graphs are regular, so using the adjacency matrix to effect the quantum walk is equivalent to using the Laplacian \cite{Wong19}. Also, Johnson graphs are vertex-transitive, so regardless of which vertex is marked, the algorithm behaves the same way. Finally, since the search Hamiltonian \eref{eq:H} is time-independent, the solution to Schr\"odinger's equation is
\begin{equation}
	\label{eq:evolution}
	\ket{\psi(t)} = \rme^{-\rmi H t} \ket{\psi(0)}.
\end{equation}

The runtime of the continuous-time quantum walk search algorithm is known for some specific cases of Johnson graphs. As introduced above, $J(n,1)$ corresponds to the complete graph $K_n$, and the search algorithm is simply the continuous-time quantum walk formulation of Grover's algorithm, which runs in $O(\sqrt{N})$ time \cite{Grover1996,FG1998a,CG2004,Wong10} (where the number of vertices \eref{eq:N} is $N = n$ in this case). Similarly, the triangular graphs $J(n,2)$ with $n \ge 4$ are strongly regular graphs, which are also known to support fast quantum search in $O(\sqrt{N})$ time for large $N$ \cite{Wong5} (where the number of vertices \eref{eq:N} is $N = n(n-1)/2$ in this case).

\begin{figure}
\begin{center}
	\includegraphics{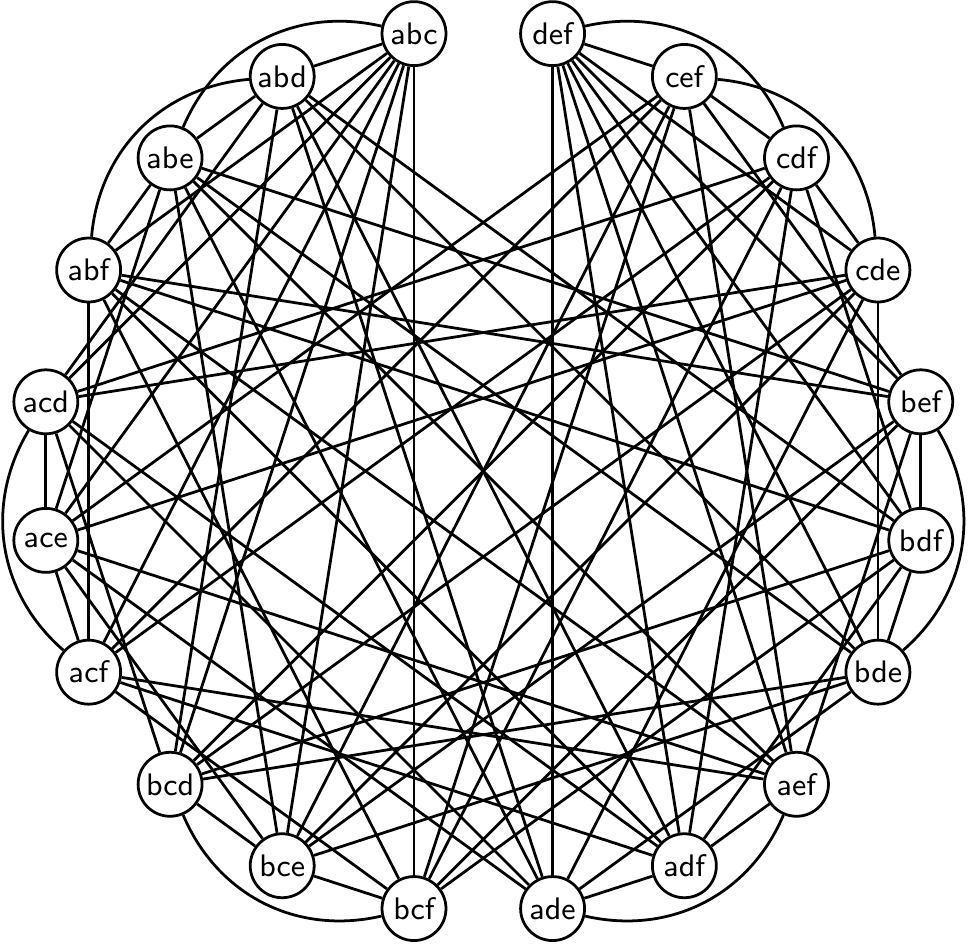}
	\caption{\label{fig:tetrahedral6} The Johnson graph $J(6,3)$, which is the $6$-tetrahedral graph.}
\end{center}
\end{figure}

Thus search on Johnson graphs with $k = 1$ and $k = 2$ is fast, achieving the full quadratic Grover speedup. In the next section, we explicitly prove that search on the next case, $k = 3$, is also fast, searching in $O(\sqrt{N})$ time, with $N = n(n-1)(n-2)/6$ the number of vertices \eref{eq:N}. In particular, $J(n,3)$ is the $n$-tetrahedral graph for $n \ge 6$ \cite{Bose1967}. For example, $J(6,3)$ is shown in \fref{fig:tetrahedral6}, and it seems to be the first illustration of a tetrahedral graph in the literature. To construct it, we take the $n = 6$ symbols to be $\{a,b,c,d,e,f\}$, and the vertices are $(k = 3)$-element subsets of the symbols. The vertices are adjacent if they differ by one symbol (or, equivalently, when they share two symbols). Also, although it is hard to tell from the figure, tetrahedral graphs are vertex-transitive, as are all Johnson graphs, so the algorithm will behave identically regardless of which vertex is marked.

The proof that search on $J(n,3)$ is fast involves reducing the evolution to a 4D subspace, which is found by taking equal superpositions of identically-evolving vertices as orthonormal basis states. Often, analyzing the energy eigensystem in this basis using degenerate perturbation theory \cite{Wong5} is sufficient to find the evolution of the algorithm \cite{Wong7,Wong9,Wong11,Wong16,Wong19}. In this case, however, the calculation is not precise enough to derive accurate dynamics. We show that a change of basis, followed by approximations as in \cite{Wong9}, does allow degenerate perturbation theory to derive accurate dynamics.

We end by outlining how this method can be applied to search on Johnson graphs $J(n,k)$ with other values of $k$, such as $k = 4, 5, 6$, etc. Each value of $k$ seems to require repeating the calculations, which is tedious and unending (since there is an infinite number of possible values of $k$). So we leave these calculations, and the case of general $k$, for further research.


\section{Search on Tetrahedral Graphs $J(n,3)$}


\subsection{4D Subspace}

\begin{figure}
\begin{center}
	\includegraphics{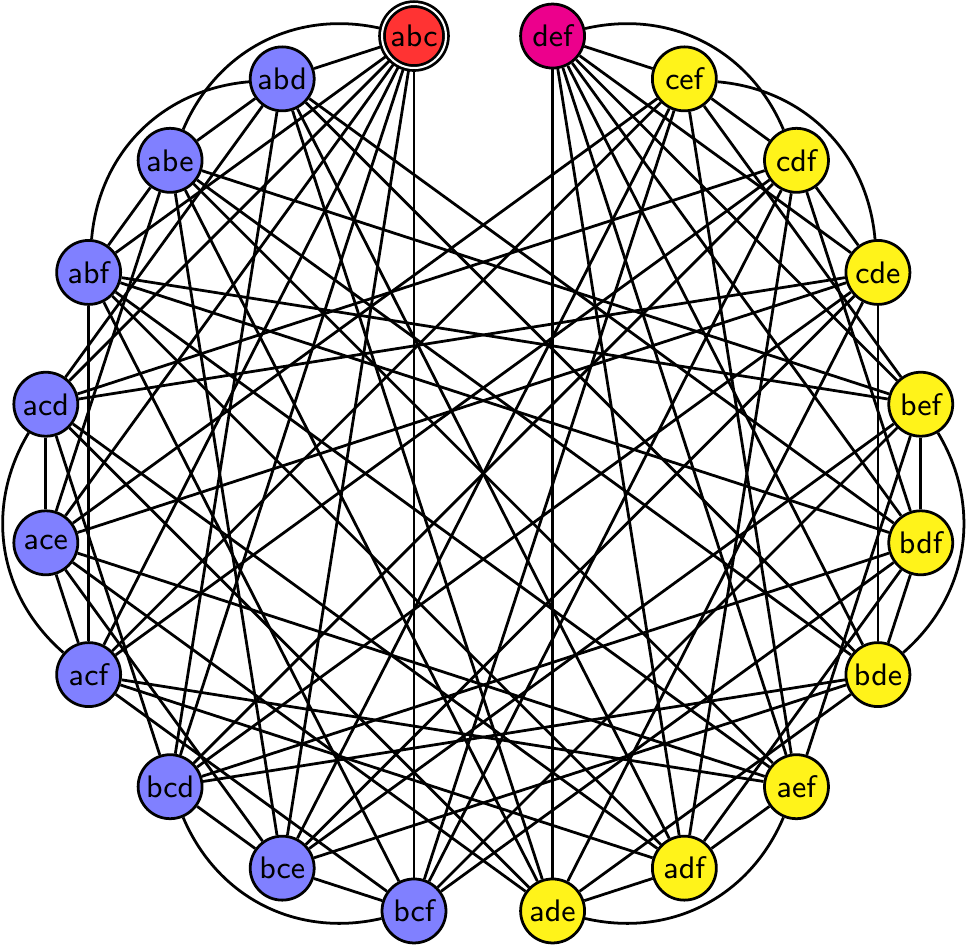}
	\caption{\label{fig:tetrahedral6_marked} The Johnson graph $J(6,3)$, which is the $6$-tetrahedral graph. Without loss of generality, a vertex is marked, as indicated by the double circle. Identically evolving vertices are identically colored.}
\end{center}
\end{figure}

We begin our analysis of search on tetrahedral graphs $J(n,3)$ by continuous-time quantum walk by noting that Johnson graphs are distance-transitive \cite{HS1993}, meaning vertices the same distance from a given vertex are all structurally identical \cite{Biggs1994}. Thus if we mark a vertex, as indicated in \fref{fig:tetrahedral6_marked} by a double circle and colored red, then the vertices adjacent (one away) from the marked vertices evolve identically. These are colored blue. Similarly, vertices two away from the marked vertex evolve identically, and they are colored yellow. Finally, vertices three away from the marked vertex, colored magenta, evolve identically (in this diagram with $n = 6$, there is just one such vertex). Let us respectively call these types $d_0$, $d_1$, $d_2$, and $d_3$, where the subscript indicates the distance from the marked vertex.

Note that tetrahedral graphs $J(n,3)$ have diameter $3$, meaning vertices are at most distance $3$ from each other. As a simple proof, consider two non-adjacent vertices $abc$ and $xyz$ of an arbitrary tetrahedral graph, with $x,y,z \not\in \{a,b,c\}$. Then one can connect them through three edges, such as $abc \sim abz \sim ayz \sim xyz$. Thus $d_0$, $d_1$, $d_2$, and $d_3$ are the only types of vertices. That is, there is no $d_4$, for example. So the system evolves in a 4D subspace. This is also the source of the constraint that $n \ge 6$; if $n < 6$, then the graph has smaller diameter.

How many vertices of each type are there? Let us denote this as $|d_i|$ and determine it for each type:
\begin{itemize}
	\item	$d_0$: Clearly, there is only one $d_0$ vertex, namely the marked vertex. That is,
		\[ |d_0| = 1. \]
		
	\item	$d_1$: Without loss of generality, say the marked vertex has symbols $abc$. Then the vertices adjacent to it have the form $xbc$, $axc$, and $abx$, where $x \not\in \{a,b,c\}$. Since each of these three forms leaves $n-3$ possibilities for $x$,
		\[ |d_1| = 3(n-3). \]
		Note this is also the degree of the graph, since it is regular.
		
	\item	$d_2$: Type $d_2$ vertices are two away from the marked vertex $abc$, so they have the form $axy$, $xby$, and $xyc$ with $x,y \not\in \{a,b,c\}$ and $x \ne y$. For each of the three forms, there are $n-3$ options for $x$ and $n-4$ options for $y$. Thus
		\[ |d_2| = \frac{3(n-3)(n-4)}{2}, \]
		where we divide by $2$ because the order of $x$ and $y$ does not matter.

	\item	$d_3$: Finally, type $d_3$ vertices are three away from the marked vertex $abc$, so they have the form $xyz$ with $x,y,z \not\in \{a,b,c\}$ and $x \ne y \ne z \ne x$. Thus there are $n-3$ options for $x$, $n-4$ options for $y$, and $n-5$ options for $z$. Thus
		\[ |d_3| = \frac{(n-3)(n-4)(n-5)}{6}, \]
		where we divide by $6 = 3!$ because the order of $x,y,z$ does not matter.
\end{itemize}
As a check, it is straightforward to show that $|d_0| + |d_1| + |d_2| + |d_3|$ equals the total number of vertices $N = n(n-1)(n-2)/6$, which comes from \eref{eq:N} with $k = 3$.

Now we take equal superpositions of each type of vertex, which together serve as an orthonormal basis of the 4D subspace:
\begin{eqnarray*}
	\ket{d_0} = \ket{w} \\
	\ket{d_1} = \frac{1}{\sqrt{|d_1|}} \sum_{d(v,w) = 1} \ket{v} = \frac{1}{\sqrt{3(n-3)}} \sum_{d(v,w) = 1} \ket{v} \\
	\ket{d_2} = \frac{1}{\sqrt{|d_2|}} \sum_{d(v,w) = 2} \ket{v} = \sqrt{\frac{2}{3(n-3)(n-4)}} \sum_{d(v,w) = 2} \ket{v} \\
	\ket{d_3} = \frac{1}{\sqrt{|d_3|}} \sum_{d(v,w) = 3} \ket{v} = \sqrt{\frac{6}{(n-3)(n-4)(n-5)}} \sum_{d(v,w) = 3} \ket{v}.
\end{eqnarray*}
In this $\{ \ket{d_0}, \ket{d_1}, \ket{d_2}, \ket{d_3} \}$ basis, the initial equal superposition state \eref{eq:s} is
\begin{eqnarray}
	\ket{s} &= \frac{1}{\sqrt{N}} \Biggl( \ket{d_0} + \sqrt{3(n-3)} \ket{d_1} + \sqrt{\frac{3(n-3)(n-4)}{2}} \ket{d_2} \nonumber \\
		&\quad\quad\quad\quad + \sqrt{\frac{(n-3)(n-4)(n-5)}{6}} \ket{d_3} \Biggr) \nonumber \\
		&= \frac{1}{\sqrt{N}} \left( \!\! \begin{array}{c}
			1 \\
			\sqrt{3(n-3)} \\
			\sqrt{\frac{3(n-3)(n-4)}{2}} \\
			\sqrt{\frac{(n-3)(n-4)(n-5)}{6}} \\
		\end{array} \!\! \right) \label{eq:s_tetrahedral} .
\end{eqnarray}
From this initial state, the system evolves by Schr\"odinger's equation with Hamiltonian given in \eref{eq:H}. To find the Hamiltonian in the 4D subspace, let us first find the adjacency matrix.

To find the adjacency matrix in the $\{ \ket{d_0}, \ket{d_1}, \ket{d_2}, \ket{d_3} \}$ basis, recall that tetrahedral graphs (and general Johnson graphs) are distance-transitive. Then the relation between each type of vertex can be summarized by its intersection array \cite{Bose1967}:
\[ \left( \! \begin{array}{cccc}
	* & 1 & 4 & 9 \\
	0 & n-2 & 2(n-4) & 3(n-6) \\
	3(n-3) & 2(n-4) & n-5 & * \\
\end{array} \!\! \right). \]
To explain this intersection array, let us start with the left column, which indicates that a $d_0$ vertex is adjacent to no other $d_0$ vertices (obviously, since there is only one marked vertex) and adjacent to $3(n-3)$ vertices of type $b$. Now for the second column, a type $d_1$ vertex is adjacent to one $d_0$ vertex, $n-2$ other $d_1$ vertices, and $2(n-4)$ type $d_2$ vertices. For the third column, a $d_2$ vertex is adjacent to four $d_1$ vertices, $2(n-4)$ other $d_2$-vertices, and $n-5$ type $d_3$ vertices. Finally, for the last column, a $d_3$ vertex is adjacent to nine $d_2$ vertices and $3(n-6)$ other $d_3$ vertices. Note that each column of the intersection array sums up to $3(n-3)$, the degree of the graph, as expected.

Using this intersection array and the number of each type of vertex, we get the adjacency matrix in the $\{ \ket{d_0}, \ket{d_1}, \ket{d_2}, \ket{d_3} \}$ basis:
\[ A = \left( \!\!\! \begin{array}{cccc}
	0 & \sqrt{3(n-3)} & 0 & 0 \\
	\sqrt{3(n-3)} & n-2 & 2\sqrt{2(n-4)} & 0 \\
	0 & 2\sqrt{2(n-4)} & 2(n-4) & 3\sqrt{n-5} \\
	0 & 0 & 3\sqrt{n-5} & 3(n-6) \\
\end{array} \!\! \right). \]
For example, the last entry of the third row is $3\sqrt{n-5}$ because a $d_2$ vertex is adjacent to $n-5$ type $d_3$ vertices, but we multiply it by $\sqrt{|d_3|/|d_2|}$ to adjust the normalization factor between the basis states $\ket{d_2}$ and $\ket{d_3}$. That is, if we denote the elements of $A$ as $A_{ij}$ with $i,j = 0,1,2,3$, then
\[ A_{ij} = ({\rm number\ of\ } d_j {\rm\ vertices\ adjacent\ to\ a\ } d_i {\rm\ vertex}) \sqrt{\frac{|d_j|}{|d_i|}}. \]

\begin{figure}
\begin{center}
	\subfloat[] {
		\includegraphics{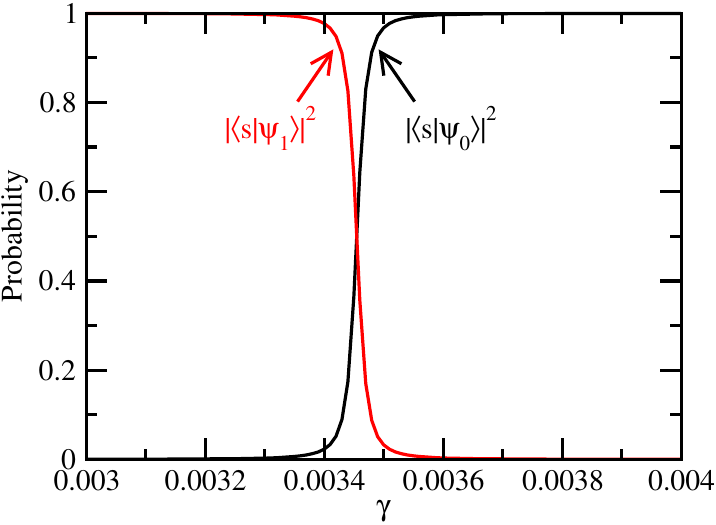}
		\label{fig:overlap_s}
	} \quad
	\subfloat[] {
		\includegraphics{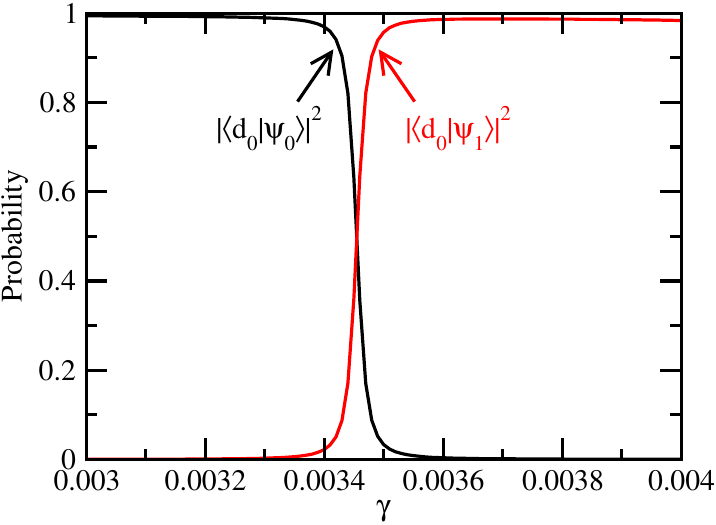}
		\label{fig:overlap_w}
	}
	\caption{Probability overlaps of (a) the initial equal superposition state $\ket{s}$ and (b) the marked vertex $\ket{d_0}$ with the eigenvectors $\ket{\psi_i}$ of the search Hamiltonian $H$ on $J(100,3)$.}
\end{center}
\end{figure}

To get the search Hamiltonian \eref{eq:H}, we simply multiply the adjacency matrix by $-\gamma$ and include the oracle $-\ketbra{w}{w} = -\ketbra{d_0}{d_0}$:
\begin{equation}
	\label{eq:H_tetrahedral}
	H = -\gamma \left( \!\!\! \begin{array}{cccc}
		\frac{1}{\gamma} & \sqrt{3(n-3)} & 0 & 0 \\
		\sqrt{3(n-3)} & n-2 & 2\sqrt{2(n-4)} & 0 \\
		0 & 2\sqrt{2(n-4)} & 2(n-4) & 3\sqrt{n-5} \\
		0 & 0 & 3\sqrt{n-5} & 3(n-6) \\
	\end{array} \!\! \right).
\end{equation}


\subsection{Numerical Simulations}

To get a sense for how the algorithm depends on the jumping rate $\gamma$, we plot in \fref{fig:overlap_s} the probability overlaps of the initial state $\ket{s}$ with the eigenvectors of $H$ for various values of $\gamma$ \cite{CG2004}. From this figure, we see that if $\gamma$ is small (towards the left side of the plot), then $\ket{s}$ is approximately equal to the first excited state $\ket{\psi_1}$ of $H$. Then we roughly start in an eigenstate of $H$, so the system only evolves by acquiring a global, unobservable phase \eref{eq:evolution}. Similarly, if $\gamma$ is large (towards the right side of the plot), then $\ket{s}$ is approximately equal to the ground state $\ket{\psi_0}$ of $H$. So again, the system roughly remains in a uniform distribution over the vertices.

When $\gamma$ takes its critical value $\gamma_c$, however, where the support of $\ket{s}$ experiences a phase transition \cite{CG2004} and is equally supported by both $\ket{\psi_0}$ and $\ket{\psi_1}$, then the evolution is nontrivial. In \fref{fig:overlap_s}, we identify that $\gamma_c \approx 0.003455$. At this critical value, \fref{fig:overlap_s} and \fref{fig:overlap_w} together indicate that the ground state and first excited state are each half in $\ket{s}$ and half in $\ket{d_0}$. That is, $\ket{\psi_{0,1}} \propto \ket{s} \pm \ket{d_0}$. This implies \cite{CG2004} that the system evolves from $\ket{s}$ to the marked vertex $\ket{d_0}$ in time $\pi/\Delta E$, where $\Delta E$ is the energy gap between the ground and first excited states.

\begin{figure}
\begin{center}
	\includegraphics{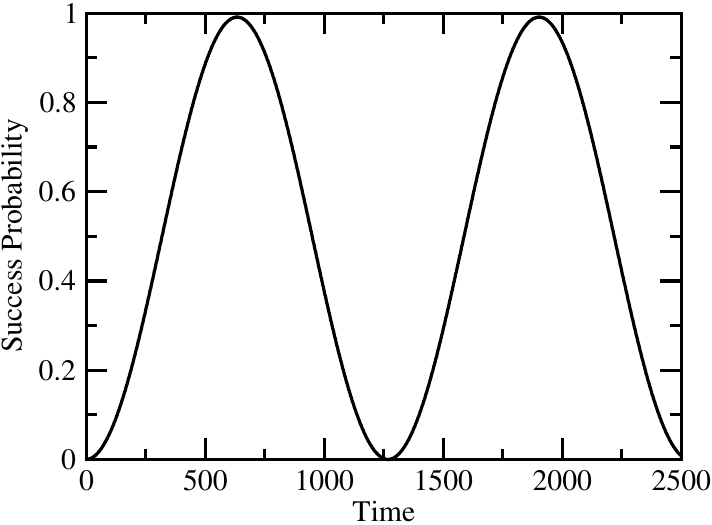}
	\caption{\label{fig:prob_time_n100} For search on $J(100,3)$, the success probability as a function of time with $\gamma = 0.003455$.}
\end{center}
\end{figure}

To demonstrate that this evolution occurs, we numerically simulate the evolution of the system $\ket{\psi(t)}$ according to \eref{eq:evolution} with $\gamma = 0.003455$ and plot in \fref{fig:prob_time_n100} the success probability $|\braket{d_0}{\psi(t)}|^2$ as the system evolves with time. We see that the system evolves to the marked vertex $\ket{d_0}$ with probability $1$ at time roughly $\pi\sqrt{N}/2 = \pi\sqrt{161700}/2 \approx 631.65$, where we used $N = n(n-1)(n-2)/6$, with $n = 100$, for the number of vertices \eref{eq:N}. This implies that the energy gap is
\[ \Delta E = \frac{2}{\sqrt{N}} \approx \frac{2\sqrt{6}}{n^{3/2}}, \]
since $N \approx n^3/6$ for large $N$.

Furthermore, using this energy gap, we can find how precisely the critical jumping rate must be determined. Say we are $\epsilon$ away from the true critical value, i.e., $\gamma = \gamma_c + \epsilon$. Then from the argument in Section VI of \cite{Wong16}, a calculation using degenerate perturbation theory \cite{Wong5} would contain a leading-order term in $\epsilon$ that scales as $\Theta(n \epsilon)$. For this term to be negligible, it must scale smaller than the energy gap: $n \epsilon = o\left( \Delta E \right)$. Solving for $\epsilon$,
\[ \epsilon = o\left( \frac{1}{n^{5/2}} \right). \]
Thus we must determine $\gamma_c$ up to $o(1/n^{5/2})$.

With this sense for how the algorithm evolves, let us analytically find the critical jumping rate $\gamma_c$ and prove that, at this value of $\gamma$, the algorithm finds the marked vertex $\ket{d_0}$ with probability $1$ at time $\pi\sqrt{N}/2 = O(\sqrt{N})$, thus achieving the full Grover speedup. We do this by finding the eigensystem of the search Hamiltonian \eref{eq:H_tetrahedral}. Unfortunately, analytically finding this eigensystem is messy, so as in \cite{Wong5,Wong7,Wong8,Wong9,Wong11,Wong16,Wong19}, we attempt to approximate it using degenerate perturbation theory. As we will see, in this case, a change of basis is required for the calculation to succeed.


\subsection{Failure of Perturbation Theory}

The idea of perturbation theory is to first find the eigensystem of a simpler matrix, then see how higher-order corrections (i.e., the perturbation) change the eigensystem. To do this, we break the search Hamiltonian \eref{eq:H_tetrahedral} into leading and higher-order terms:
\[ H = H^{(0)} + H^{(1)} + H^{(2)} + \dots, \]
where
\[
	H^{(0)} = -\gamma \left( \!\! \begin{array}{cccc}
		\frac{1}{\gamma} & 0 & 0 & 0 \\
		0 & n & 0 & 0 \\
		0 & 0 & 2n & 0 \\
		0 & 0 & 0 & 3n \\
	\end{array} \!\! \right),
\]
\[
	H^{(1)} = -\gamma \left( \!\!\! \begin{array}{cccc}
		0 & \sqrt{3n} & 0 & 0 \\
		\sqrt{3n} & 0 & 2\sqrt{2n} & 0 \\
		0 & 2\sqrt{2n} & 0 & 3\sqrt{n} \\
		0 & 0 & 3\sqrt{n} & 0 \\
	\end{array} \!\! \right),
\]
and so forth. It is now easy to find the eigensystem of the leading-order Hamiltonian $H^{(0)}$; its eigenvectors are $\ket{d_0}$, $\ket{d_1}$, $\ket{d_2}$, and $\ket{d_3}$ with respective eigenvalues $-1$, $-\gamma n$, $-2\gamma n$, and $-3\gamma n$. If these eigenvalues are not degenerate, perturbation theory \cite{Wong5} implies that adding higher-order corrections will not significantly change the eigenvectors. Then $\ket{d_3}$ remains an approximate eigenvector of the system. Since the initial state $\ket{s} \approx \ket{d_3}$ for large $N$ \eref{eq:s_tetrahedral}, this implies that the system roughly starts in an eigenvector of $H$, so it fails to evolve apart from a global, unobservable phase \eref{eq:evolution}.

When the leading-order eigenvectors are degenerate, however, then the perturbation can cause a dramatic change \cite{Wong5}. Say $\gamma = 1/3n$ so that $\ket{d_0}$ and $\ket{d_3}$ are degenerate eigenvectors of $H^{(0)}$, with both of them having eigenvalue $-1$. Then the perturbation causes two linear combinations of $\ket{d_0}$ and $\ket{d_3}$ to be eigenvectors of the perturbed system. This choice of $\gamma = 1/3n$ yields an estimate for the critical $\gamma$. Often, such a calculation is precise enough \cite{Wong7,Wong9,Wong11,Wong16,Wong19}, but in this case, it is not. If $n = 100$, then this estimate for $\gamma_c$ yields $1/300 \approx 0.0033$, which from \fref{fig:overlap_s} is not close enough to the actual value of $0.003455$. In the last section, we showed that $\gamma_c$ must be known up to terms $o(1/n^{5/2})$; while we have the leading-order term $1/3n$, we need a correction that scales as $\Theta(1/n^2)$. That is, the next correction beyond that would scale as $\Theta(1/n^3)$, which we can ignore.

\begin{figure}
\begin{center}
	\subfloat[] {
		\includegraphics{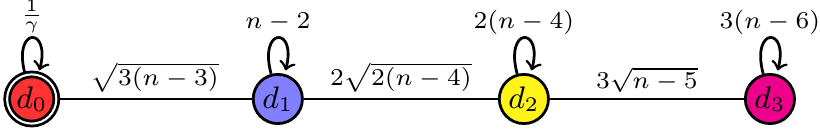}
		\label{fig:diagram_H}
	}

	\subfloat[] {
		\includegraphics{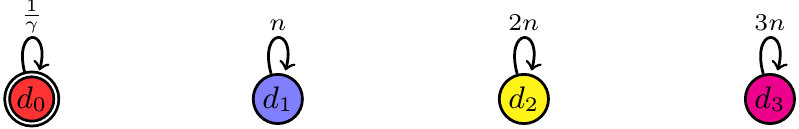}
		\label{fig:diagram_H0}
	}

	\subfloat[] {
		\includegraphics{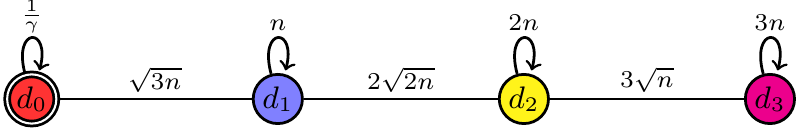}
		\label{fig:diagram_H0H1}
	}

	\caption{Apart from a factor of $-\gamma$, (a) the Hamiltonian for search on the tetrahedral graph $J(n,3)$. (b) The terms scaling as $n$. (c) The terms up scaling greater than or equal to $\sqrt{n}$.}
\end{center}
\end{figure}

The perturbative calculation is also not accurate enough for another reason, which we illustrate by expressing the calculation diagrammatically \cite{Wong8}. \Fref{fig:diagram_H} expresses the full search Hamiltonian $H$, and \fref{fig:diagram_H0} expresses its leading-order terms $H^{(0)}$. When $\gamma = 1/3n$, $\ket{d_0}$ and $\ket{d_3}$ are degenerate in \fref{fig:diagram_H0} since they have self-loops of the same weight. With the perturbation $H^{(1)}$, we get \fref{fig:diagram_H0H1}, which restores the missing edges. But since there is no edge connecting $\ket{d_0}$ and $\ket{d_3}$, these states do not mix; the perturbed eigenstates roughly remain $\ket{d_0}$ and $\ket{d_3}$, not a combination of the two, and so probability does not flow between the vertices.

One might try to improve the precision of the perturbative calculation by letting $H^{(0)}$ contain higher-order terms, like for the ``simplex of complete graphs'' in \cite{Wong7,Wong9,Wong16}. In our case, however, including both terms that scale as $\Theta(n)$ and $\Theta(\sqrt{n})$ in $H^{(0)}$ yields a matrix, equivalent to \fref{fig:diagram_H0H1}, whose eigensystem is still too messy to determine.


\subsection{Change of Basis and Success of Perturbation Theory}

We can circumvent this failure of degenerate perturbation theory by changing the basis. Rather than using superpositions of identically evolving vertices $\{ \ket{d_0}, \ket{d_1}, \ket{d_2}, \ket{d_3} \}$ as basis states, we use
\[ \ket{d_0} \]
\begin{eqnarray*}
       	\ket{r}
		&= \frac{1}{\sqrt{N-1}} \sum_{v \ne w} \ket{v} \\
		&= \sqrt{\frac{18}{n^2+2}} \left( \ket{d_1} + \sqrt{\frac{n-4}{2}} \ket{d_2} + \sqrt{\frac{(n-4)(n-5)}{18}} \ket{d_3} \right)
\end{eqnarray*}
\[ \ket{r'} = \sqrt{\frac{9}{n+4}} \left( -\sqrt{\frac{n-5}{9}} \ket{d_2} + \ket{d_3} \right) \]
\[ \ket{r''} = \frac{9\sqrt{2}}{\sqrt{(n^2+2)(n+4)}} \left( \frac{(n+4)\sqrt{n-4}}{9\sqrt{2}} \ket{d_1} - \ket{d_2} - \frac{\sqrt{n-5}}{3} \ket{d_3} \right). \]
This basis resembles the one used in \cite{Wong5} for strongly regular graphs, which includes triangular graphs $J(n,2)$.

To convert the search Hamiltonian \eref{eq:H_tetrahedral} from the $\{ \ket{d_0}, \ket{d_1}, \ket{d_2}, \ket{d_3} \}$ basis to the $\{ \ket{d_0}, \ket{r}, \ket{r'}, \ket{r''} \}$ basis, we compute $T^{-1} H T$, where
\[ T = \left( \begin{array}{cccc}
	\ket{d_0} & \ket{r} & \ket{r'} & \ket{r''} \\
\end{array} \right). \]
Doing this, we get $H$ in the new basis, which we call $H'$:
\[ \arraycolsep=1pt \footnotesize H' \! = \! -\gamma \! \left( \! \begin{array}{cccc}
	\frac{1}{\gamma} & \frac{3 \sqrt{6(n-3)}}{\sqrt{n^2+2}} & 0 & \frac{\sqrt{3(n-3)(n-4)(n+4)}}{\sqrt{n^2+2}} \\
	\frac{3 \sqrt{6(n-3)}}{\sqrt{n^2+2}} & \frac{3 (n^3-3n^2+2n-12)}{n^2+2} & 0 & \frac{-3 \sqrt{2(n-4)(n+4)}}{n^2+2} \\
	0 & 0 & \frac{2n^2-9n-32}{n+4} & \frac{-2 \sqrt{2(n-5)(n^2+2)}}{n+4} \\
	\frac{\sqrt{3(n-3)(n-4)(n+4)}}{\sqrt{n^2+2}} & \frac{-3 \sqrt{2(n-4)(n+4)}}{n^2+2} & \frac{-2 \sqrt{2(n-5)(n^2+2)}}{n+4} & \frac{n^4+2n^3-42n^2+22n-16}{(n+4)(n^2+2)}
\end{array} \! \right) \!\! . \]

\begin{figure}
\begin{center}
	\subfloat[] {
		\includegraphics{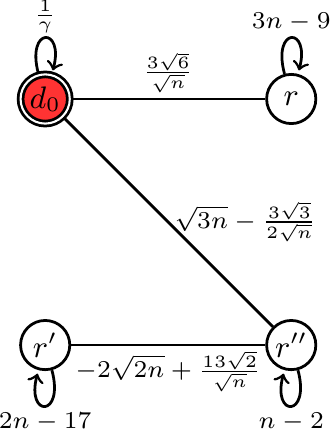}
		\label{fig:diagram_H0H1_basis}
	} \quad \quad \quad \quad
	\subfloat[] {
		\includegraphics{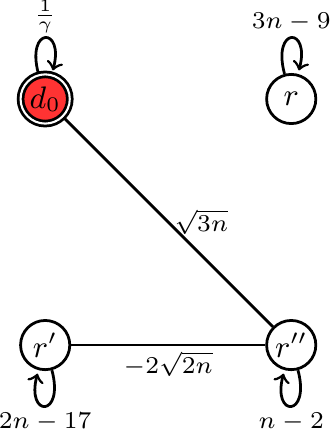}
		\label{fig:diagram_H0_basis}
	}

	\caption{Apart from a factor of $-\gamma$, (a) $H'$ up to terms $o(1/\sqrt{n})$, and (b) terms up to $o(1)$.}
\end{center}
\end{figure}

Now we use degenerate perturbation theory to analyze $H'$. To get a better sense for how each term scales with $n$, we write $H'$ up to $o(1/\sqrt{n})$:
\[ H' \approx -\gamma \left( \!\! \begin{array}{cccc}
	\frac{1}{\gamma} & \frac{3\sqrt{6}}{\sqrt{n}} & 0 & \sqrt{3n} - \frac{3\sqrt{3}}{2\sqrt{n}} \\
	\frac{3\sqrt{6}}{\sqrt{n}} & 3n-9 & 0 & 0 \\
	0 & 0 & 2n-17 & -2 \sqrt{2n} + \frac{13\sqrt{2}}{\sqrt{n}} \\
	\sqrt{3n} - \frac{3\sqrt{3}}{2\sqrt{n}} & 0 & -2 \sqrt{2n} + \frac{13\sqrt{2}}{\sqrt{n}}  & n-2 \\
\end{array} \!\! \right). \]
This is diagrammatically represented in \fref{fig:diagram_H0H1_basis}, and it reveals an important distinction from the $\{ \ket{d_0}, \ket{d_1}, \ket{d_2}, \ket{d_3} \}$ basis in \fref{fig:diagram_H}. Note that $\ket{s} \approx \ket{r}$, and now there is an edge connecting $\ket{r}$ and $\ket{d_0}$. Thus for the perturbative calculation, we take the simpler matrix to be
\[ H'^{(0)} = -\gamma \left( \!\! \begin{array}{cccc}
	\frac{1}{\gamma} & 0 & 0 & \sqrt{3n} \\
	0 & 3n-9 & 0 & 0 \\
	0 & 0 & 2n-17 & -2 \sqrt{2n} \\
	\sqrt{3n} & 0 & -2 \sqrt{2n} & n-2 \\
\end{array} \!\! \right). \]
Then we include the perturbation
\[ H'^{(1)} = -\gamma \left( \!\! \begin{array}{cccc}
	0 & \frac{3\sqrt{6}}{\sqrt{n}} & 0 & -\frac{3\sqrt{3}}{2\sqrt{n}} \\
	\frac{3\sqrt{6}}{\sqrt{n}} & 0 & 0 & 0 \\
	0 & 0 & 0 & \frac{13\sqrt{2}}{\sqrt{n}} \\
	-\frac{3\sqrt{3}}{2\sqrt{n}} & 0 & \frac{13\sqrt{2}}{\sqrt{n}} & 0 \\
\end{array} \!\! \right), \]
which restores the edge between $\ket{r}$ and $\ket{d_0}$. Diagrammatically, we first find the eigensystem of $H^{(0)}$ depicted in \fref{fig:diagram_H0_basis}, then the addition of the perturbation $H^{(1)}$ gives \fref{fig:diagram_H0H1_basis}.

So we begin by finding the eigensystem of $H^{(0)}$. Clearly, one eigenvector is $\ket{r}$ with its corresponding eigenvalue $E_r = -\gamma(3n-9)$:
\[ \ket{r}, \quad E_r = -\gamma(3n-9). \]
The remaining three eigenvectors are combinations of $\ket{d_0}$, $\ket{r'}$ and $\ket{r''}$, and the Hamiltonian corresponding to them is
\[ H'^{(0)}_{d_0,r',r''} = -\gamma \left( \begin{array}{cccc}
	\frac{1}{\gamma} & 0 & \sqrt{3n} \\
	0 & 2n-17 & -2 \sqrt{2n} \\
	\sqrt{3n} & -2 \sqrt{2n} & n-2 \\
\end{array} \right). \]
The eigenvectors of this are messy, so we follow the method of Section A.3 and A.4 of \cite{Wong9} to approximate them. The eigenvalues $\lambda$ of $H'^{(0)}_{d_0,r',r''}$ satisfy the characteristic equation
\begin{eqnarray*}
	 -\lambda^3 &- (3 \gamma n - 19\gamma + 1) \lambda^2 + \gamma \left[ 19 - 34\gamma - 2 \gamma n^2 + n(32\gamma - 3) \right] \lambda \\
	&+ \gamma^2 \left[ -34 + n(29 - 51\gamma) + n^2(-2 + 6\gamma) \right] = 0.
\end{eqnarray*}
When $\gamma$ is
\begin{equation}
	\label{eq:gamma_c}
	\gamma_c = \frac{1}{3n} + \frac{7}{6n^2},
\end{equation}
then $E_r$ and one of these eigenvalues, which we will call $\lambda_u$, both equal
\[ -1 - \frac{1}{2n} + O\left( \frac{1}{n^2} \right), \]
so $E_r$ and $\lambda_u$ are degenerate up to terms of $O(1/n^2)$. So we have determined the critical $\gamma$ up to $o(1/n^{5/2})$, which is the required precision we previously determined. For example, with $n = 100$, this yields $\gamma_c = 1/(3\cdot100) + 7/(6\cdot100^2) \approx 0.00345$, which from \fref{fig:overlap_s} is accurate enough.

Let us find the eigenvector of $H'^{(0)}_{d_0,r',r''}$ corresponding to $\lambda_u$, and let us call it $\ket{u}$. To find it, we solve $H'^{(0)}_{d_0,r',r''} \ket{u} = \lambda_u \ket{u}$:
\[ -\gamma \left( \!\! \begin{array}{ccc}
	\frac{1}{\gamma} & 0 & \sqrt{3n} \\
	0 & 2n-17 & -2 \sqrt{2n} \\
	\sqrt{3n} & -2 \sqrt{2n} & n-2 \\
\end{array} \!\! \right) \left( \!\! \begin{array}{c}
	u_{d_0} \\
	u_{r'} \\
	u_{r''} \\
\end{array} \!\! \right) = \lambda_u \left( \!\! \begin{array}{c}
	u_{d_0} \\
	u_{r'} \\
	u_{r''} \\
\end{array} \!\! \right). \]
Using the first row,
\[ -u_{d_0} - \gamma\sqrt{3n} u_{r''} = \lambda_u u_{d_0} \quad \Rightarrow \quad u_{r''} = -\frac{1+\lambda_u}{\gamma\sqrt{3n}} u_{d_0}. \]
Using the second row,
\begin{eqnarray*}
	&(2n-17)u_{r'} - 2\sqrt{2n}u_{r''} = \frac{-\lambda_u}{\gamma} u_{r'} \\ &\quad \Rightarrow \quad u_{r'} = \frac{2\sqrt{n}}{2n-17+\frac{\lambda_u}{\gamma}} u_{r''} = -\frac{2\sqrt{n}}{2n-17+\frac{\lambda_u}{\gamma}} \frac{1+\lambda_u}{\gamma\sqrt{3n}} u_{d_0} .
\end{eqnarray*}
So we have
\[ \ket{u} = u_{d_0} \left( \!\! \begin{array}{c}
	1 \\
	-\frac{2\sqrt{n}}{2n-17+\frac{\lambda_u}{\gamma}} \frac{1+\lambda_u}{\gamma\sqrt{3n}} \\
	-\frac{1+\lambda_u}{\gamma\sqrt{3n}} \\
\end{array} \!\! \right), \]
where $u_{d_0}$ is selected to normalize the state. Note that substituting $\lambda_u = -1 - 1/2n$, we find that $\ket{u} \approx \ket{d_0}$ for large $N$.

We have that $\ket{r}$ and $\ket{u}$ are approximately degenerate eigenvectors of $H^{(0)}$. Now let us include the perturbation $H^{(1)}$. With this, two linear combinations of $\ket{r}$ and $\ket{u}$,
\[ \alpha_r \ket{r} + \alpha_u \ket{u}, \]
become eigenstates of the perturbed system. To find the coefficients, we solve
\[ \left( \!\! \begin{array}{cc}
	H'_{rr} & H'_{ru} \\
	H'_{ur} & H'_{uu} \\
\end{array} \!\! \right) \left( \!\! \begin{array}{c}
	\alpha_r \\
	\alpha_u \\
\end{array} \!\! \right) = E \left( \!\! \begin{array}{c}
	\alpha_r \\
	\alpha_u \\
\end{array} \!\! \right), \]
where $H'_{ru} = \langle r | H'^{(0)} + H'^{(1)} | u \rangle$, etc. Evaluating the matrix components, we get
\[ \left( \!\! \begin{array}{cc}
	-1 - \frac{1}{2n} + O(1/n^2) & \frac{-\sqrt{6}}{n^{3/2}} + O(1/n^{5/2}) \\
	\frac{-\sqrt{6}}{n^{3/2}} + O(1/n^{5/2}) & -1 - \frac{1}{2n} - O(1/n^2) \\
\end{array} \!\! \right) \left( \!\! \begin{array}{c}
	\alpha_r \\
	\alpha_u \\
\end{array} \!\! \right) = E \left( \!\! \begin{array}{c}
	\alpha_r \\
	\alpha_u \\
\end{array} \!\! \right). \]
Solving this up to terms $O(1/n^{5/2})$, we get perturbed eigenstates and eigenvalues
\[ \frac{1}{\sqrt{2}} \left( \ket{r} \pm \ket{u} \right), \quad E_\pm = -1 \mp \frac{\sqrt{6}}{n^{3/2}}. \]
Thus we have proved that the system evolves from the initial equal superposition state $\ket{s} \approx \ket{r}$ to the marked vertex $\ket{d_0} \approx \ket{u}$ in time
\[ \frac{\pi}{\Delta E} = \frac{\pi}{2} \sqrt{\frac{n^3}{6}} \approx \frac{\pi}{2}\sqrt{N} \]
for large $N$, which is the full quantum quadratic speedup.

\begin{figure}
\begin{center}
	\includegraphics{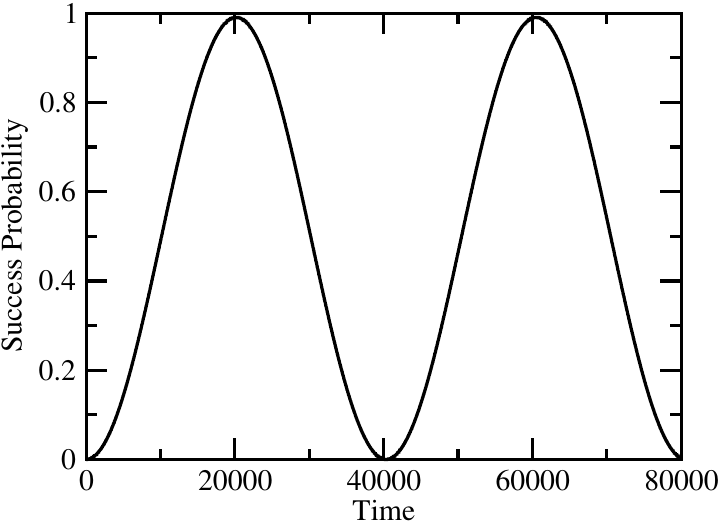}
	\caption{\label{fig:prob_time_n1000} For search on $J(1000,3)$, the success probability as a function of time with $\gamma = \gamma_c = 1/3n + 7/6n^2$.}
\end{center}
\end{figure}

To illustrate these results, the evolution of the search algorithm for $J(1000,3)$ is shown in \fref{fig:prob_time_n1000} using the critical $\gamma$ we derived \eref{eq:gamma_c}. The success probability approaches $1$ at time $\pi\sqrt{N}/2 \approx 20248.5$, as expected.

\section{Search on General Johnson Graphs $J(n,k)$}

We end by outlining how the above method for analyzing search on tetrahedral graphs $J(n,3)$ can be generalized to arbitrary Johnson graphs $J(n,k)$ with fixed $k$.

As previously noted, Johnson graphs are distance-transitive, so vertices of equal distance from a given vertex are structurally identical. As in the last section, we denote the types of vertices $d_0, d_1, \dots, d_k$, where the subscript indicates the distance from the marked vertex. Then the number of each type of vertex is
\begin{eqnarray*}
	|d_i| 
		&= \frac{\binom{k}{i} (n-k) (n-k-1) \dots (n-k-(i-1))}{i!} = \frac{\binom{k}{i} (n-k)!}{(n-k-i)! i!} \\
		&= \binom{k}{i} \binom{n-k}{i}.
\end{eqnarray*}
Taking equal superpositions of each type of vertex
\[ \ket{d_i} = \frac{1}{\sqrt{|d_i|}} \sum_{d(v,w) = i} \ket{v} = \frac{1}{\sqrt{\binom{k}{i} \binom{n-k}{i}}} \sum_{d(v,w) = i} \ket{v}, \]
then $\{ \ket{d_0}, \ket{d_1}, \dots, \ket{d_k} \}$ forms an orthonormal basis for the $(k+1)$-dimensional subspace of the evolution.

\begin{figure}
\begin{center}
	\includegraphics{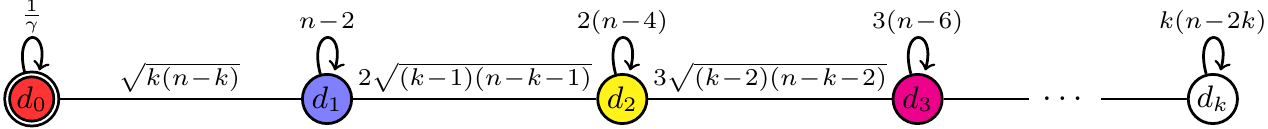}
	\caption{\label{fig:diagram_H_Johnson} Apart from a factor of $-\gamma$, the Hamiltonian for search on the Johnson graph $J(n,k)$.}
\end{center}
\end{figure}

Now let us express the algorithm in this basis. The system starts in an equal superposition over the vertices:
\[ \ket{s} = \frac{1}{\sqrt{N}} \sum_{i = 0}^k \sqrt{|d_i|} \ket{d_i} = \frac{1}{\sqrt{N}} \sum_{i = 0}^k \sqrt{\binom{k}{i} \binom{n-k}{i}} \ket{d_i}. \]
Then it evolves by Schr\"odinger's equation with Hamiltonian \eref{eq:H}. To find the Hamiltonian in the $(k+1)$-dimensional basis, we use the number of each type of vertex along with the intersection array (Section 8 of \cite{HS1993})
\[ \arraycolsep=2pt \footnotesize \left( \! \begin{array}{cccccc}
	* & 1 & 4 & 9 & \dots & k^2 \\
	0 & n-2 & 2(n-4) & 3(n-6) & \dots & k(n-2k) \\
	k(n-k) & (k-1)(n-k-1) & (k-2)(n-k-2) & (k-3)(n-k-3) & \dots & * \\
\end{array} \!\! \right) \]
to determine the adjacency matrix of $J(n,k)$:
\[ \hspace{-0.9in} \arraycolsep=1pt \footnotesize A = \left( \!\! \begin{array}{cccccc}
	0 & \sqrt{k(n-k)} & 0 & 0 & \cdots & 0 \\
	\sqrt{k(n-k)} & n-2 & 2\sqrt{(k-1)(n-k-1)} & 0 & \cdots & 0 \\
	0 & 2\sqrt{(k-1)(n-k-1)} & 2(n-4) & 3\sqrt{(k-2)(n-k-2)} & \cdots & 0 \\
	0 & 0 & 3\sqrt{(k-2)(n-k-2)} & 3(n-6) & \cdots & 0 \\
	\vdots & \vdots & \vdots & \vdots & \ddots & \vdots \\
	0 & 0 & 0 & 0 & \cdots & k(n-2k) \\
\end{array} \! \right) \!. \]
Then the search Hamiltonian \eref{eq:H} is simply
\[ \hspace{-1in} \arraycolsep=1pt \footnotesize H \! = \! -\gamma \! \left( \!\!\! \begin{array}{cccccc}
	\frac{1}{\gamma} & \sqrt{k(n-k)} & 0 & 0 & \cdots & 0 \\
	\sqrt{k(n-k)} & n-2 & 2\sqrt{(k-1)(n-k-1)} & 0 & \cdots & 0 \\
	0 & 2\sqrt{(k-1)(n-k-1)} & 2(n-4) & 3\sqrt{(k-2)(n-k-2)} & \cdots & 0 \\
	0 & 0 & 3\sqrt{(k-2)(n-k-2)} & 3(n-6) & \cdots & 0 \\
	\vdots & \vdots & \vdots & \vdots & \ddots & \vdots \\
	0 & 0 & 0 & 0 & \cdots & k(n-2k) \\
\end{array} \!\! \right) \!\! . \]
This search Hamiltonian is diagrammatically represented in \fref{fig:diagram_H_Johnson}, and from this we identify that degenerate perturbation theory will have the same problems for $J(n,k)$ as it did for tetrahedral graphs $J(n,3)$.

To rescue the calculation, we can again change the basis, with two of the $k+1$ basis vectors being 
\[ \ket{d_0} \]
and
\[ \ket{r} = \frac{1}{\sqrt{N-1}} \sum_{v \ne w} \ket{v}. \]
In the new basis, one would perform the perturbative calculation and show that the system approximately evolves from $\ket{s} \approx \ket{r}$ to some vector $\ket{u} \approx \ket{d_0}$. This outlines how our method can be applied to search on $J(n,k)$ with fixed $k$, where one would need to work through the calculations for each $k$ (e.g., $k = 4, 5, 6, \dots$). We leave such calculations, and whether it can be done for general $k$ (i.e., when $k$ is a variable and not a number), as questions for further research.


\section{Conclusion}

We have identified that previously known results about fast continuous-time quantum walk search on the complete graph and triangular graphs are the $k = 1$ and $k = 2$ instances of Johnson graphs $J(n,k)$. We expanded these results, showing that search on tetrahedral graphs $J(n,3)$ is also fast, and we used degenerate perturbation theory with a change of basis to do so. This method can be used for other values of $k$, but we leave those calculations, and the general problem of search with arbitrary $k$, as topics for further investigation.


\ack
This work was supported by the European Union Seventh Framework Programme (FP7/2007-2013) under the QALGO (Grant Agreement No.~600700) project, and the ERC Advanced Grant MQC.


\section*{References}
\bibliographystyle{iop}
\bibliography{refs}

\end{document}